\documentclass[12pt,a4paper,epsf]{article}
\usepackage{graphics}
\usepackage{amssymb,amsmath}
\usepackage[dvips]{lscape,graphicx}
\usepackage{cite}

\newcommand{\ct}{\cite}
\newcommand{\lb}{\label}

\newcommand{\bc}{\begin{center}}
\newcommand{\ec}{\end{center}}
\newcommand{\bd}{\begin{displaymath}}
\newcommand{\ed}{\end{displaymath}}
\newcommand{\be}{\begin{equation}}
\newcommand{\ee}{\end{equation}}
\newcommand{\ba}{\begin{array}}
\newcommand{\ea}{\end{array}}
\newcommand{\bea}{\begin{eqnarray}}
\newcommand{\eea}{\end{eqnarray}}
\newcommand{\bt}{\begin{tabular}}
\newcommand{\et}{\end{tabular}}

\newcommand{\ov}{\overline}
\newcommand{\bp}{\begin{picture}}
\newcommand{\ep}{\end{picture}}
\newcommand{\bfi}{\begin{figure}}
\newcommand{\efi}{\end{figure}}

\begin{document}

\hyphenation{ }

\title{\Large\bf Are Preons Dyons ?\\ Naturalness of Three Generations}

\author{\large C.R.~Das ${}^{1}$ \footnote{\large\, crdas@imsc.res.in}\,\,
and Larisa Laperashvili ${}^{1,\, 2}$ \footnote{\large\, laper@itep.ru, 
laper@imsc.res.in}\\[5mm]
\itshape{${}^{1}$ The Institute of Mathematical Sciences, Chennai, India}
\\[0mm]
\itshape{${}^{2}$ The Institute of Theoretical and Experimental Physics, Moscow,
 Russia}}

\date{}
\maketitle

\vspace{1cm}













\maketitle



\begin{abstract}

In the present paper a model of preons and their composites is constructed
in the framework of the superstring-inspired `flipped' $E_6\times \widetilde{E_6}$
gauge symmetry group which reveals a generalized dual symmetry.
Here $E_6$ and $\widetilde{E_6}$ are non-dual and dual sectors of the theory 
with hyper-electric $g$ and hyper-magnetic $\tilde g$ charges, respectively.
We follow the old idea 
by J.~Pati presuming that preons are dyons, which in our model are confined
by hyper-magnetic strings -- composite ${\rm\bf N}=1$ supersymmetric non-Abelian flux 
tubes created by the condensation of spreons near the Planck scale. 
Considering preons belonging to the 27-plet of $E_6$, we have
investigated the breakdown of $E_6$ (and $\widetilde{E_6}$) near the Planck 
scale into the $SU(6)\times U(1)$ (and $\widetilde{SU(6)}\times 
\widetilde{U(1)}$) gauge group and shown that the six types of 
strings having fluxes $\Phi_n=n\Phi_0$ $(n=\pm 1,\pm 2,\pm 3)$ produce three 
(and only three) generations of composite quark-leptons and bosons. The model 
predicts the existence of the Family replicated gauge group $[E_6]^3$ near the 
Planck scale. The runnings of the corresponding fine structure constants are 
investigated, and the critical coupling constants at the two phase transition 
points near the Planck scale are calculated. The model explains the hierarchies
of masses in the Standard Model naturally. The compactification in a 
space-time with five dimensions and its influence on form-factors of composite 
objects are also briefly discussed.

\end{abstract}

\clearpage\newpage

\section{Introduction}

During the last twenty years Grand Unified Theories (GUTs) have been inspired 
by the ultimate
theory of superstrings \ct{1,2,3}, which gives the possibility of unifying all 
fundamental interactions including gravity. It was shown in Ref.~\ct{4} 
that strings are free of gravitational and Yang-Mills anomalies if
the gauge group is $SO(32)$ or $E_8\times E_8$. The ``heterotic" 
$E_8\times E'_8$ superstring theory was suggested in Ref.~\ct{2} as a more 
realistic candidate for unification. This ten-dimensional Yang-Mills 
theory can undergo a spontaneous compactification for which the $E_8$ group is 
broken to $E_6$ in four dimensions. The group $E'_8$ remains unbroken 
and plays the role of a hidden sector, which provides the spontaneous 
breakdown of SUGRA and a soft SUSY breaking mechanism.
As a result, superstring theory leads to the SUSY-GUT scenario in four 
dimensions.

In the present paper we develop a new preonic model of composite quark-leptons 
and bosons starting with the `flipped' $E_6\times \widetilde{E_6}$ 
supersymmetric group. We show that a dual sector of the theory described
by the group $\widetilde{E_6}$ is broken in our world near 
the Planck scale $M_{Pl}\approx 1.22\cdot 10^{19}$ GeV.
The breakdown of the dual sector gives in our model a very specific type of 
``horizontal symmetry" predicting only three generations of the Standard Model.
 
We consider that preons are dyons confined by hyper-magnetic strings,
which are created by the condensation of spreons near the Planck scale.

Pati first \ct{5} suggested the use of the strong $U(1)$ magnetic force
to bind preons-dyons and form composite objects.
This idea was developed in Refs.~\ct{6,6a,6b} (see also the review \ct{7}) and 
is extended in our model in the light of recent investigations of 
composite non-Abelian flux tubes in SQCD \ct{8a,8b,8c,8d,8e,8f} (see also 
Refs.~\ct{8p,8q,8r} and the review \ct{8s}).
A very interesting solution for a QCD string with quantized flux was found in 
Ref.~\ct{9a}.

A supersymmetric preon model based on the (non-flipped) 
$E_6\times {E'}_6$ gauge group was investigated in Ref.~\ct{9} 
with different properties.

A preonic model of composite quarks and leptons in higher space-time 
dimensions was constructed in Ref.~\ct{10}. Their results are utilized in 
our present model of preons. In general, composite models were reviewed
in Refs.~\ct{10a,10b,10c,10d,10e,10x,10f,10g,10k}.

The paper is organized as follows. In Sect.~2 we have presented a feasible
chain of flipped models, which may exist between $M_{GUT}\sim 10^{16}$ GeV
and $M_{Pl}\sim 10^{19}$ GeV ended by the flipped $E_6$ unification, and have 
considered the content of a 27-plet of the flipped $E_6$. Sect.~3 contains an 
${\rm\bf N}=1$ supersymmetric description of $E_6$-theory in five space-time dimensions.
Sect.~4 presents a new preonic model of quark-leptons and bosons, in which it is 
suggested that preons are dyons confined by hyper-magnetic 
strings. The breakdown of $E_6$ and $\widetilde{E_6}$ symmetries and the 
condensation of spreons near the Planck scale are discussed in this Section. 
It is shown that the composite non-Abelian flux tubes constructed from preonic strings 
are very thin ($\sim 10^{-18}$ GeV$^{-1}$) and have 
an enormously large tension ($\sim 10^{38}$ GeV$^2$). 
The naturalness of three generations (families) of quark-leptons and bosons 
is also demonstrated in Sect.~4 to be a consequence of the quantized flux of 
tubes considered in our theory.   
In Sect.~5 the existence of the Family replicated gauge group $[E_6]^3$ is
discussed, and the breakdown of $[E_6]^3$ and $[\widetilde{E_6}]^3$ near the Planck 
scale is analysed. Sect.~6 is devoted to the solution of the problem of mass
hierarchies established in the Standard Model. Also form factors indicating
the compositeness of quark-leptons are discussed in this section. The results of 
the present investigation are summarized in our Conclusions (Sect.~7).

\section{Flipped $E_6$-unification of gauge interactions }

In Ref.~\ct{11}, where we consider that only `flipped' $SU(5)$ unifies $SU(3)_C$ 
and $SU(2)_W$ of the Standard Model (SM) at the GUT scale $M_{GUT}\sim 10^{16}$ 
GeV, we gave an explanation of the discrepancy between the unification 
scale $M_{GUT}$ and string scale $M_{str}\sim 10^{18}$ GeV. We have assumed 
that there exists a chain of extra intermediate symmetries between $M_{GUT}$ 
and $M_{Pl}$:
\be
      SU(5)\times U(1)_X \to  SU(5)\times U(1)_{Z1}
\times U(1)_{X1} \to $$ $$ SO(10) \times U(1)_{X1} \to  E_6. \lb{1}
\ee
Amongst the many articles devoted to the $E_6$ unification (see, for example,
\ct{11aa,11aaa,11a,11b} and references therein) we have chosen only one example of the 
flipped $E_6$ unification from Ref.~\ct{11}. We have considered such Higgs 
boson contents of the $SU(5)$ and $SO(10)$ gauge groups which give the 
final flipped $E_6$ unification at the scale $\sim 10^{18}$ GeV and a decreased 
running of the inverse gauge coupling constant $\alpha^{-1}$ near the Planck 
scale. Such an example, presented in Fig.~\ref{f1}, suits the purposes of our
new model of preons.    
Here and below we consider flipped models in which $SU(5)$ contains 
Higgs bosons $h,h^{\rm\bf c}$ and $H,H^{\rm\bf c}$ belonging to $5_h + \bar 5_h$ and 
$10_H + \ov{10}_H$ representations of $SU(5)$, respectively, also a
24-dimensional adjoint Higgs field $A$ and Higgs bosons belonging
to additional higher representations. Correspondingly, the flipped $SO(10)$ 
(coming at the superGUT scale $M_{SG}$) contains  $10_h + \ov {10}_h$ and 
$45_H + \ov{45}_H$, a 45-dimensional adjoint $A$ and higher representations of 
Higgs bosons. As was shown in \ct{11}, such Higgs boson contents lead 
to the flipped $E_6$ final unification at the supersuperGUT scale 
$M_{SSG}\sim 10^{18}$ GeV. 

Fig.~\ref{f1} presents an example of the running of the inverse gauge coupling 
constants $\alpha_i^{-1}(\mu)$ ($\mu$ is the energy scale) for 
$i=1,2,3,X,Z,X1,Z1,5,10$. It was shown that at the scale $\mu = M_{GUT}$ the 
flipped $SU(5)$ undergoes breakdown to the supersymmetric (MSSM) 
$SU(3)_C\times SU(2)_L\times U(1)_Z\times U(1)_X$ gauge group, 
which is the supersymmetric extension of the MSSM
originating at the seesaw scale $M_{SS}\approx 10^{11}$ GeV, where heavy 
right-handed neutrinos appear. A singlet Higgs field $S$ provides the 
following breakdown to the SM (see \ct{12}):
\be SU(3)_C\times SU(2)_L\times U(1)_Z\times U(1)_X \to
    SU(3)_C\times SU(2)_L\times U(1)_Y.                            \lb{2} 
\ee
Supersymmetry extends the conventional SM beyond the scale $M_{SUSY}$ 
(in our example $M_{SUSY} = 10$ TeV).

The final $E_6$ unification assumes the existence of three 27-plets of $E_6$ 
containing three generations of quarks and leptons including the right-handed 
neutrinos $N_i^{\rm\bf c}$ (here $i=1,2,3$ is the generation index). Quarks and 
leptons of the fundamental 27 representation decompose under the 
$SU(5)\times U(1)_X$ subgroup as follows:
\be
        27 \to (10,1) + \left(\bar 5, -3\right) +  \left(\bar 5,2\right) +
          (5,-2) + (1,5) + (1,0).                             \lb{3}
\ee
The first and second quantities in the brackets of Eq.~(\ref{3}) correspond to 
the $SU(5)$ representation and $U(1)_X$ charge, respectively. 
We consider charges $Q_X$ and $Q_Z$ in units of ${1}/{\sqrt{40}}$ 
and $\sqrt{3/5}$, respectively, using the assignments: $Q_X=X$ and $Q_Z=Z$ 
(see \ct{11} for details).

The conventional SM family, which contains doublets of left-handed quarks 
$Q$ and leptons $L$, right-handed up and down quarks $u^{\rm\bf c}$, $d^{\rm\bf c}$, 
also $e^{\rm\bf c}$,  
is assigned to the $(10,1) + (\bar 5,-3) + (1,5)$ representations 
of the flipped $SU(5)\times U(1)_X$, along with a right-handed neutrino $N^{\rm\bf c}$. 
These representations decompose under
\be
SU(5)\times U(1)_X \to SU(3)_C\times SU(2)_L\times U(1)_Z\times U(1)_X,      
                                                     \lb{4a}
\ee
and such a decomposition gives the following content:

\bea
       (10,1) \to Q = &\left(\begin{array}{c}u\\ 
                                          d \end{array}\right) &\sim 
                         \left(3,2,\frac 16,1\right),\nonumber\\
&d^{\rm\bf c} &\sim \left(\bar3,1,-\frac 23,1\right),\nonumber\\
&N^{\rm\bf c} &\sim \left(1,1,1,1\right).       \lb{4}\\
\left(\bar 5,-3\right) \to &u^{\rm\bf c}&\sim \left(\bar 3,1,\frac 13,-3\right),\nonumber\\
L = &\left(\begin{array}{c}e\\ 
                                             \nu \end{array}\right) &\sim 
                         \left(1,2,-\frac 12,-3\right),               \lb{5}\\
(1,5) \to &e^{\rm\bf c} &\sim \left(1,1,1,5\right).\lb{6}
\eea

The remaining representations in Eq.~(\ref{3}) decompose as follows:

\bea
        (5,-2) \to& D&\sim \left(3,1,-\frac 13,-2\right),\nonumber\\
                   h = &\left(\begin{array}{c}h^+\\ 
                                               h^0 \end{array}\right) &\sim 
                         \left(1,2,\frac 12,-2\right).
                                                              \lb{7}\\
    \left(\bar 5,2\right) \to &D^{\rm\bf c} &\sim \left(\bar 3,1,\frac 13,2\right),\nonumber\\
                     h^{\rm\bf c} = &\left(\begin{array}{c}h^0\\ 
                                               h^- \end{array}\right) &\sim 
                         \left(1,2,-\frac 12,2\right).              \lb{8}
\eea
The light Higgs doublets are accompanied by coloured Higgs triplets $D,D^{\rm\bf c}$.

The singlet field $S$ is represented by (1,0):
\be
       (1,0) \to S \sim (1,1,2,2).               \lb{9}
\ee
It is necessary to notice that the flipping of our $SU(5)$:
\be
       d^{\rm\bf c} \leftrightarrow u^{\rm\bf c},\quad 
N^{\rm\bf c}\leftrightarrow e^{\rm\bf c}, \lb{10}
\ee
distinguishes this group of symmetry from the standard Georgi-Glashow $SU(5)$ 
\ct{13}.

The multiplets (\ref{4})--(\ref{6}) fit in the 16 spinorial representation 
of $SO(10)$:
\be
       F(16) = F(10,1) + F\left(\bar 5,-3\right) + F(1,5).     \lb{11}
\ee
Higgs chiral superfields belong to the 10 representation of $SO(10)$:
\be 
       h(10) = h(5,-2) + h^{\rm\bf c}\left(\bar 5,2\right).            \lb{12}
\ee
In Section 5 we predict that in the interval of energy 
$M_{SSG}\le \mu \le M_{crit}$ (see Fig.~\ref{f2}) the Family replicated gauge group of 
symmetry $[E_6]^3$ originates at the scale $M_{FR}$: $M_{SSG}< M_{FR}< M_{crit}$. 

\section{Supersymmetric $E_6$ content  in five space-time 
dimensions}

Taking into account the role of compactification \ct{14} 
we start, as in Ref.~\ct{10},
with ${\rm\bf N}=1$ supersymmetric gauge theory in $5D$ dimensions 
with a local symmetry gauge group $G$, which is equal to the 
flipped $E_6$ in our case. The matter fields transform according to one of 
the irreducible 
representations of $G$: for example, we have the 27-plet
of $E_6$ given by Eqs.~(\ref{4})--(\ref{9}).
 
The supermultiplet ${\cal V} = (A^M, \lambda^{\alpha}, \Sigma, X^a)$ in 
$5D$-dimensional space (with $M=0,1,2,3,4$ space-time indices) contains 
a vector field:
\be
            A^M = A^{MJ}T^J,     \lb{13}                                      
\ee
and a real scalar field:
\be
                  \Sigma = \Sigma^JT^J,             \lb{14}
\ee
where $J$ runs over the $E_6$ group index values and $T^J$ are generators 
of the $E_6$ algebra.

Two gauginos fields:
\be
            \lambda^{\alpha} = \lambda^{\alpha J}T^J,        \lb{15}
\ee
form a decuplet under the $R$-symmetry group $SU(2)_R$ (with $\alpha=1,2$).

Auxiliary fields:
\be     
                  X^a = X^{aJ}T^J                      \lb{16}
\ee
form a triplet of $SU(2)_R$ (with $a=1,2,3$).

After compactification, these fields are combined into the ${\rm\bf N}=1$ $4D$ fields:
$$
   {\mbox{vector supermultiplet}}\quad V=\left(A^{\mu}, \lambda^1, X^3\right)
$$
(here $\mu=0,1,2,3$ are the space-time indices), and
$$
 {\mbox{a chiral supermultiplet}}\quad \Phi=\left(\Sigma+iA^4, \lambda^2, 
X^1+iX^2\right).
$$
The matter fields are contained in the hypermultiplet:
$$
       {\cal H} = (h^{\alpha}, \Psi, F^{\alpha}),
$$
where $h^{\alpha}$ is a doublet of $SU(2)_R$, the Dirac fermion field 
$\Psi=(\psi_1, \psi_2^+)^T$
is an $SU(2)_R$ singlet and the auxiliary fields $F^{\alpha}$ also form an  
$SU(2)_R$ doublet.

Then we have two ${\rm\bf N}=1$ $4D$ chiral multiplets:
$$
              H = \left(h^1, \psi_1, F^1\right)\quad {\rm and}\quad H^{\rm\bf c} = 
                                 \left(h^2, \psi_2, F^2\right)  
$$ 
transforming according to the representations $R$ and anti-$R$ of the gauge group 
$G=E_6$, respectively.
Then the $5D$-dimensional $E_6$-symmetric action (see \ct{15}) is given as 
follows:
\be
       S = \int d^5x\int d^4\theta\left[H^{\rm\bf c}e^VH^{{\rm\bf c}+} + H^+e^VH\right] + 
        \int d^5x\int d^2\theta\left[H^{\rm\bf c} \left(\partial_4 - 
\frac 1{\sqrt 2}\Phi\right)H 
                                + h.c.\right].   \lb{17}           
\ee
This theory is anomaly-free \ct{10,15}.
The compactification results are given in Ref.~\ct{10}, and we summarize them
in Appendix A.

\section{Supersymmetric  $E_6\times \widetilde {E_6}$ preonic model
of composite quark-leptons and bosons }

Why do three generations exist in Nature? We suggest an 
explanation considering a new preonic model of composite SM particles.
The model starts from the supersymmetric flipped $E_6\times \widetilde {E_6}$ 
gauge symmetry group for preons, where  $E_6$ and $\widetilde {E_6}$ 
correspond to non-dual (with hyper-electric charge $g$) and dual (with 
hyper-magnetic charge $\tilde g$) sectors of the theory, respectively. 
We assume that preons are dyons confined by hyper-magnetic strings in the 
region of energy $\mu \le \widetilde{M}_{crit} \sim M_{Pl}$ (see Fig.~\ref{f2}).

\subsection{Preons are dyons bound by hyper-magnetic strings}

Considering the ${\rm\bf N}=1$ supersymmetric flipped $E_6\times \widetilde {E_6}$
gauge theory for preons in $4D$-dimensional space-time, we assume that preons 
$P$ (antipreons $P^{\rm\bf c}$) are 
dyons with charges $ng$ and $m\tilde g$ ($-ng$ and $-m\tilde g$), residing
in the $4D$ hypermultiplets ${\cal P} = (P,P^{\rm\bf c})$ and ${\cal \tilde P} = 
(\tilde P, {\tilde P}^{\rm\bf c})$. Here ``$\tilde P$" designates spreons, but not 
the belonging to $\widetilde{E_6}$.

The dual sector $\widetilde{E_6}$ is broken in our world to some group $\widetilde{G}$,
where preons and spreons transform under the hyper-electric gauge group $E_6$ 
and hyper-magnetic gauge group $\widetilde{G}$ according to their fundamental 
representations:
\be
   P,\tilde P\sim (27,N), \quad  P^{\rm\bf c},\tilde P^{\rm\bf c}\sim 
\left(\ov {27},\bar N\right),
                                         \lb{19}
\ee
where $N$ is the $N$-plet of $\widetilde{G}$ group.
We also consider scalar preons and spreons as singlets of $E_6$:
\be
   P_s,\tilde P_s\sim (1,N), \quad  P_s^{\rm\bf c},\tilde P_s^{\rm\bf c}\sim 
\left(1,\bar N\right),
                                         \lb{20}
 \ee  
which are actually necessary for the entire set of composite 
quark-leptons and bosons \ct{10}. 

The hyper-magnetic interaction is assumed to be responsible for the formation
of $E_6$ fermions and bosons at the compositeness scale $\Lambda_s$.
The main idea of the present investigation is the assumption that
preons-dyons are confined by hyper-magnetic supersymmetric 
non-Abelian flux tubes, which are a generalization of the well-known 
Abelian ANO-strings \ct{17,18} for the case of the supersymmetric 
non-Abelian theory developed in Refs.~\ct{8a,8b,8c,8d,8e,8f,8p,8q,8r,8s}.
As a result, in the limit of infinitely narrow flux tubes (strings), we have 
the following bound states:

\begin{itemize}
\item[{\bf i.}] quark-leptons (fermions belonging to the $E_6$ fundamental representation):
\bea
            Q^a &\sim& P^{aA}(y) \left[{\cal P}\exp\left(i\tilde g
\int_x^y\widetilde{A}_{\mu}dx^{\mu}\right)\right]_A^B
                         (P_s^{\rm\bf c})_B(x) \sim 27,
                                               \lb{21}
\\
          \bar  Q_a &\sim& (P_s^{\rm\bf c})^A(y) 
\left[{\cal P}\exp\left(i\tilde g\int_x^y\widetilde{A}_{\mu}dx^{\mu}\right)\right]_A^B
                         P_{aB}(x) \sim \ov {27},
                                               \lb{22}
\eea     
where $a\in 27$-plet of $E_6$, $A, B\in N$-plet of $\widetilde{G}$, and 
$\widetilde{A}_{\mu}(x)$ are dual hyper-gluons belonging to the adjoint 
representation of $\widetilde{G}$;

\item[{\bf ii.}] ``mesons" (hyper-gluons and hyper-Higgses of $E_6$):
\bea
            M^a_b &\sim& P^{aA}(y) \left[{\cal P}\exp\left(i\tilde g
\int_x^y\widetilde{A}_{\mu}dx^{\mu}\right)\right]_A^B
                         (P^{\rm\bf c})_{bB}(x)\nonumber\\ 
&&\sim 1+78+650\quad {\rm of}\,\,E_6,
                                               \lb{23}
\\
            S &\sim& (P_s)^A(y) \left[{\cal P}\exp\left(i\tilde g
\int_x^y\widetilde{A}_{\mu}dx^{\mu}\right)\right]_A^B
                         (P_{s}^{\rm\bf c})_B(x) \sim 1,
                                               \lb{24}
\\
          \bar S &\sim& (P_s^{\rm\bf c})^A(y) 
\left[{\cal P}\exp\left(i\tilde g\int_x^y\widetilde{A}_{\mu}dx^{\mu}\right)\right]_A^B
                         (P_s)_B(x) \sim 1;
                                               \lb{25}
\eea   

\item[{\bf iii.}] ``baryons",

 for $\widetilde{G}$-triplet we have (see Section 5):
\bea
             D_1 &\sim& \epsilon_{ABC}P^{aA'}(z)P^{bB'}(y)P^{cC'}(x)
\left[{\cal P}\exp\left(i\tilde g\int^z_X\widetilde{A}_{\mu}dx^{\mu}\right)\right]_{A'}^A
\times
 \nonumber\\
&&    \left[{\cal P}\exp\left(i\tilde g\int^y_X\widetilde{A}_{\mu}dx^{\mu}\right)\right]_{B'}^B
 \left[{\cal P}\exp\left(i\tilde g\int^x_X\widetilde{A}_{\mu}dx^{\mu}\right)\right]_{C'}^C,      \lb{26}
\\
             D_2 &\sim& \epsilon_{ABC}P^{aA'}(z)P^{bB'}(y)(P_s)^{C'}(x)
 \left[{\cal P}\exp\left(i\tilde g\int^z_X\widetilde{A}_{\mu}dx^{\mu}\right)\right]_{A'}^A
\times
\nonumber\\
&&   \left[{\cal P}\exp\left(i\tilde g\int^y_X\widetilde{A}_{\mu}dx^{\mu}\right)\right]_{B'}^B
 \left[{\cal P}\exp\left(i\tilde g\int^x_X\widetilde{A}_{\mu}dx^{\mu}\right)\right]_{C'}^C,      \lb{27}
\eea 
and their conjugate particles.
\end{itemize}
The bound states (\ref{21})--(\ref{27}) are shown
in Fig.~\ref{f3} as unclosed strings $(a)$ and ``baryonic'' configurations $(b)$.
It is easy to generalize Eqs.~(\ref{21})--({\ref{27}) for the case of 
string constructions of superpartners -- squark-sleptons, hyper-gluinos 
and hyper-higgsinos. Closed strings -- gravitons -- are presented 
in Fig.~\ref{f3}(c).
All these bound states belong to $E_6$ representations and they in 
fact form ${\rm\bf N}=1$ $4D$ superfields.

\subsection{The breakdown of  $E_6$  and  $\widetilde{E_6}$ groups at the Planck scale} 

As was suggested in Ref.~\ct{16}, $E_6$ can be broken by
Higgses belonging to the 78-dimensional representation of $E_6$:
\be
               E_6 \to SU(6)\times SU(2) \to SU(6)\times U(1),
                                                      \lb{28}
\ee
where $SU(6)\times U(1)$ is the largest relevant invariance group of the 78.

Fig.~\ref{f2} shows two points $A$ and $B$ near the Planck scale. The point 
$A$ corresponds to the breakdown of $E_6$ 
according to the chain (\ref{28}).
The group $E_6$ is broken in the region of energies $\mu \ge M_{crit}$ 
producing hyper-electric strings between preons.  
The point $B$ in Fig.~\ref{f2} indicates the scale 
$\widetilde{M}_{crit}$ corresponding  to the breakdown 
$\widetilde {E_6} \to \widetilde {SU(6)}\times \widetilde {U(1)}$.
At the point $B$ hyper-magnetic strings are produced and 
exist in the region of 
energies $\mu \le \widetilde{M}_{crit}$ confining hyper-magnetic charges of preons. 
As a result, in the 
region  $\mu \le M_{crit}$ we see quark-leptons with charges $ng$ $(n\in Z)$, 
but in the region  $\mu \ge \widetilde{M}_{crit}$ monopolic ``quark-leptons" 
-- particles with dual charges $m\tilde g$ $(m\in Z)$ -- may exist.
Since $\widetilde{M}_{crit} > M_{Pl}$, monopoles are absent in our world.

In the region of energies $M_{crit}\le \mu \le \widetilde{M}_{crit}$ around the 
Planck scale, both hyper-electric and hyper-magnetic strings come to play: 
preons, quarks and monopolic ``quarks" are totally confined 
giving heavy neutral particles with mass $M\sim M_{Pl}$, but closed strings -- 
gravitons -- survive there.

\subsection{ Condensation of spreons near the Planck scale,  and ${\rm\bf N}=1$ 
supersymmetric non-Abelian flux tubes confined  preons}

If the $SU(6)\times U(1)$ symmetry group works near the Planck scale, 
then we deal just with the theory of non-Abelian flux tubes in ${\rm\bf N}=1$
SQCD, which was developed recently in Refs.~\ct{8a,8b,8c,8d,8e,8f,8p,8q,8r,8s}.

We assume the condensation of spreons at the Planck scale.
One can combine the $Z_6$ center of $SU(6)$ with the elements 
$exp(i\pi )\in U(1)$ to get topologically stable string solutions possessing 
both windings, in $SU(6)$ and $U(1)$. Henceforth we assume the existence of a dual 
sector of the theory described by $\widetilde {SU(6)}\times \widetilde {U(1)}$,
which is responsible for hyper-magnetic fluxes. Then, 
according to the results obtained in Refs.~\ct{8a,8b,8c,8d,8e,8f,8p,8q,8r,8s}, we have 
a nontrivial homotopy group:
\be 
            \pi_1\left(\frac{SU(6)\times U(1)}{Z_6}\right) \neq 0, \lb{29}
\ee
and flux lines form topologically non-trivial $Z_6$ strings.

Besides $SU(6)$ and $U(1)$ gauge bosons, the model contains six scalar fields 
charged with respect to $U(1)$ which belong to the 6-plet of $SU(6)$.
Considering scalar fields of spreons  
\be
        \tilde P=\left\{ \phi^{aA}\right\}, \lb{30}
\ee
which have indices $a$ of  $SU(6)$ and $A$ of $\widetilde {SU(6)}$ 
fundamental multiplets, we construct a condensation of spreons in the vacuum:
\be
       {\tilde P}_{vac} = \left\langle\tilde P^{aA}\right\rangle = 
v\cdot{\rm diag}(1,1,..,1),\quad
                               a,A=1,..,6. \lb{31}
\ee
The vacuum expectation value (VEV) $v$ is given in Refs.~\ct{8a} as 
\be
              v =\sqrt \xi \gg \Lambda_4,           \lb{32} 
\ee
where $\xi$ is the Fayet-Iliopoulos $D$-term parameter in the ${\rm\bf N}=1$ 
supersymmetric theory and $\Lambda_4$ is its 4-dimensional scale.
In our case:
\be
             v\sim M_{Pl}\sim 10^{19}\,\,{\rm GeV},   \lb{33}    
\ee
where spreons are condensed.

Non-trivial topology (\ref{29}) amounts to the winding of elements of the matrix 
(\ref{30}), and we obtain string solutions:
\be
 {\tilde P}_{string} = v\cdot{\rm diag}\left(e^{i\alpha(x)},e^{i\alpha(x)},..,1,1\right),
                          \quad {\rm where}\quad x\to \infty. \lb{34}
\ee
Three types of string moduli space (see Ref.~\ct{8b}):
\be
   \frac{SU(6)}{SU(5)\times U(1)},\quad 
\frac{SU(6)}{SU(4)\times SU(2) \times U(1)}\quad {\rm and} \quad
\frac{SU(6)}{SU(3)\times SU(3) \times U(1)}
                                                        \lb{35}
\ee
give us solutions for three types of $Z_6$-flux tubes which form 
a non-Abelian analog of ANO-strings \ct{17,18}.

Assuming  the existence of a preon $P$ and antipreon $P^{\rm\bf c}$ 
at the ends of strings with hyper-magnetic 
charges $n\tilde g$ and $-n\tilde g$, respectively, we obtain six 
types of string having their fluxes $\Phi_n$ quantized according 
to the $Z_6$ center group of $SU(6)$:
\be
      \Phi_n = n\Phi_0, \quad n=\pm 1,\pm 2,\pm 3.
                                                       \lb{36}          
\ee
Indeed, $Z_6$ has six group elements:
\be
        Z_6 = \left\{\left.\exp\left(2\pi \frac n6 i\right)\right|
 n\,\,{\rm mod}\,\,6\right\}. 
                                   \lb{37}   
\ee
So far as  $n$ is given modulo 6, the fluxes of tubes corresponding 
to the solutions with $n=4,5$ are equal to the fluxes (\ref{36}) with
$n=-2,-1$, respectively, (see also Refs.~\ct{8e}).

The string tensions of these non-Abelian flux tubes were also calculated in
Refs.~\ct{8a}. The minimal tension is:
\be
               T_0 = 2\pi \xi,                  \lb{38}                  
\ee
which in our preonic model (see Eq.~(\ref{32})) is equal to:
\be
               T_0 = 2\pi v^2\sim 10^{38}\,\,{\rm GeV}^2.
                                                          \lb{39}
\ee
Such an enormously large tension
means that preonic strings have almost infinitely small 
$\alpha' \to 0$, where $\alpha'= 1/(2\pi T_0)$ is the slope of trajectories 
in string theory \ct{1}. Six types of preonic tubes give us three types of 
preonic $k$-strings having the following tensions:
\be            
                    T_k =kT_0,\quad {\rm where}\quad k=1,2,3. \lb{39a}
\ee
If the Fayet-Iliopoulos term $\xi$ vanishes in the supersymmetric theory
of preons, then the spreon condensate vanishes too, and the theory is in the 
Coulomb phase. But for a non-vanishing $\xi$ the spreons develop their VEV, 
and the theory is in the Higgs phase. Then hyper-magnetic charges of preons
and antipreons 
are confined by six strings which are oriented in opposite directions. For
this reason, six strings have only three different tensions (\ref{39a}). 

Also preonic strings are extremely thin. Indeed, the thickness of 
the flux tube depends on the mass of the dual gauge boson $\widetilde{A}_{\mu}$
acquired in the confinement phase:
\be
                m_V =gv.                         \lb{40}
\ee
As it is shown below, in the region of energies $AB$ near the Planck scale
where spreons are condensed (see Fig.~\ref{f2}) we have:
\be
  \alpha=\frac{g^2}{4\pi}\approx 1,\quad g\approx 2\sqrt {\pi}\approx 3.5,
                                                 \lb{41}
\ee
and the thickness of preonic strings given by the radius $R_{str}$ of 
the flux tubes is very small:
\be
      R_{str}\sim \frac 1{m_V} \sim \frac 1{gv}\sim 10^{-18}
                                          \,\,{\rm GeV}^{-1}. \lb{42}
\ee
Such infinitely narrow non-Abelian supersymmetric flux tubes remind us of 
the superstrings in Superstring theory.
Having in our preonic model  supersymmetric strings with $\alpha'\to 0$ we
obtain, according to the description \ct{1}, only massless ground states:
spin 1/2 fermions (quarks and leptons), spin 1 hyper-gluons and spin 2 
massless graviton, as well as their superpartners. The excited states 
belonging to these strings are not realized in our world as they are 
very massive: they have mass $M > M_{Pl}$. 

\section{Three generations of quark-leptons and bosons}

The hyper-flavor ``horizontal" symmetry was suggested first in 
Refs.~\ct{16a,16b} (see also interesting discussions of this problem in 
Refs.~\ct{10,16c,16d}).
The previous Section gives a demonstration of a very specific 
type of ``horizontal symmetry" (which resembles the idea of 
Refs.~\ct{19a,19b,19c}): three, and only three, generations of 
fermions and bosons present in the superstring-inspired flipped
$E_6$ theory, and also in each step given by Fig.~\ref{f1} up to
the Standard Model. This number ``3" is explained by the existence of just three  
values of the hyper-magnetic flux (\ref{36}) which bind the hyper-magnetic 
charges of preons-dyons. At the ends of these preonic strings there are placed 
hyper-magnetic charges $\pm\tilde g_0$, or $\pm 2\tilde g_0$, or 
$\pm 3{\tilde g}_0$, where ${\tilde g}_0$ is the minimal hyper-magnetic charge.
Then all the bound states (\ref{4})--(\ref{9}) form three generations -- three
27-plets of $E_6$ corresponding to the three different tube fluxes.
We also obtain three types of gauge boson $A_{\mu}^i$ ($i=1,2,3$
is the generation index) belonging to the 
$27\times \ov {27} = 1 + 78 + 650$ representations of $E_6$. 
Fig.~\ref{f4} illustrates the formation of such 
hyper-gluons (Fig.~\ref{f4}(a)) and also hyper-Higgses (Fig.~\ref{f4}(b)).

Such a description
predicts the Family replicated gauge group of symmetry $[E_6]^3$ 
for quark-leptons and bosons, which works near the Planck scale. 
Here the number of families is equal to the number of generations $N_g=3$. 
We assume that the $[E_6]^3$ symmetry holds 
in the region of energies $M_{FR} \le \mu \le M_{crit}$, where the 
scales $M_{FR}$ and $M_{crit}$ are indicated in Fig.~\ref{f2} by the points $D$ 
and $A$, respectively.
 
The symmetry breakdown $[E_6]^3\to E_6$ at the scale 
$M_{FR}$ is provided by several Higgses.  An analogous mechanism 
is described in Refs.~\ct{19,20,21} (see also reviews \ct{22,23}). Indeed,
in the Family replicated gauge group we have three types of gauge bosons 
$A_{\mu}^i$, which produce linear combinations:
\be
   A_{\mu,diag }^{(i)} = C_1^{(i)} A_{\mu}^{(1st\,\,fam.)} +    
C_2^{(i)} A_{\mu}^{(2nd\,\,fam.)} + C_3^{(i)} A_{\mu}^{(3rd\,\,fam.)},
\quad i=1,2,3.
                                                         \lb{42a}
\ee
The combination: 
\be
   A_{\mu,diag } = \frac 1{\sqrt 3}\left(A_{\mu}^{(1st\,\,fam.)} +    
    A_{\mu}^{(2nd\,\,fam.)} + A_{\mu}^{(3rd\,\,fam.)}\right)
                                                         \lb{42b}
\ee
is the massless one, which corresponds to the 78 adjoint representation 
of hyper-gluons of $E_6$. Two other combinations are massive and exist 
only in the region of energies $\mu\ge M_{FR}$. 

The assumption that only $[E_6]^3$ exists in Nature may be not valid:
we are not sure that the Family replicated gauge groups $[SO(10)]^3$ \ct{24},
$[SU(5)]^3$ \ct{25,26,27,28}, $[SMG\times U(1)_{(B-L)}]^3$ \ct{19,20,21,22}, or
$[SMG]^3$ \ct{29,30,31,32} do not survive at lower energies $\mu < M_{FR}$.
Here we have used the following notation: $SMG$ is the Standard Model group
$SU(3)_C\times SU(2)_L\times U(1)_Y$.

The values of $\alpha^{-1}(M_{crit})$ and $\alpha^{-1}(\widetilde{M}_{crit})$
can be calculated by the method developed in Refs.~\ct{33,34}. If the $SU(N)$ group
is broken by Abelian scalar particles belonging to its Cartan subalgebra $U(1)^{N-1}$, 
then the $SU(N)$ critical coupling constant $\alpha_N^{crit}$ is given 
by the following expression in the one-loop approximation (see \ct{33}):
\be
  \alpha_N^{crit}\approx \frac N2 \sqrt{\frac{N+1}{N-1}}\alpha_{U(1)}^{crit},
                                                               \lb{43} 
\ee
where $\alpha_{U(1)}^{crit}$ is the critical coupling constant for the
Abelian $U(1)$ theory. 

Here it is necessary to distinguish the $E_6$ gauge symmetry group for 
preons from the $E_6$ for quark-leptons.  
The points $A$ and $B$ of Fig.~\ref{f2} respectively correspond to the
breakdowns of $[E_6]^3$ and $[\widetilde{E}_6]^3$ in the region $AB$ 
of spreon condensation. In that region we have the breakdown \ct{16} of the preon 
(one family) $E_6$ (or $\widetilde{E}_6$):  
$$   E_6 \to SU(6)\times U(1) \quad \left({\rm or} \quad 
                  \widetilde{E}_6 \to \widetilde {SU(6)}\times 
                                        \widetilde {U(1)}\right),
$$
and the scale  $M_{crit}$ (or $\widetilde{M}_{crit}$)
is the scale of breaking.

In our preonic model the group $SU(6)$ is broken by condensed Abelian 
scalar spreons belonging to the Cartan subalgebra $U(1)^5$ 
(compare with the results of Ref.~\ct{34}). 
Then for the one family of preons we have:
\be
           \alpha_6^{crit}\approx 3.55\alpha_{U(1)}^{crit},   \lb{44}
\ee
according to Eq.~(\ref{43}) for $N=6$.

The behaviour of the effective fine structure constants in the vicinity of 
the phase transition point ``Coulomb-confinement" was investigated in the 
compact lattice $U(1)$ theory by Monte Carlo methods \ct{35,36}. The following 
result was obtained:
\be 
           \alpha_{lat.U(1)}^{crit}\approx 0.20\pm 0.015,    
\quad {\tilde \alpha}_{lat.U(1)}^{crit}\approx 1.25\pm 0.010.    
                                                                 \lb{45}
\ee
The calculation of the critical coupling constants in the Higgs scalar 
monopole (dyon) model of dual $U(1)$ theory \ct{33,37,37a} gave the following results:
\be 
\alpha_{U(1)}^{crit}\approx 0.21, \quad
{\tilde \alpha}_{U(1)}^{crit}\approx 1.20  \quad - \quad 
{\mbox{in the Higgs monopole model} }                
\lb{46}
\ee and
\be 
\alpha_{U(1)}^{crit}\approx 0.19, \quad
{\tilde \alpha}_{U(1)}^{crit}\approx 1.29   \quad - \quad {\mbox{in the Higgs dyon model}}
                          \lb{46a}
\ee
According to Eqs.~(\ref{45})--(\ref{46a}), the condensation of spreons leads 
to the following critical constants:
\be
            \alpha_6^{crit}\approx 3.55\cdot 0.2 \approx 0.71,
\quad     {\tilde \alpha}_6^{crit} = \left(\alpha_6^{crit}\right)^{-1}\approx 1.41. 
                                                   \lb{47} 
\ee
For the point $C$ shown in Fig.~\ref{f2} we have:
\be
\alpha_6^{(one\,\,fam.)}(M_{Pl})=1. \lb{fam}
\ee 
This result for one family of preons confirms the estimate (\ref{41}).
Here we have used the Dirac relation for non-Abelian theories (see the explanation
in Ref.~\ct{34}):
\be
    g\tilde g = 4\pi n, \quad n\in Z, \quad \alpha \tilde \alpha = 1.      \lb{48}
\ee
From the phase transition result (\ref{47}) for one family of preons,
we obtain the following inverse critical coupling constant for the 
quark-lepton $[E_6]^3$ at the phase transition point $A$ (see reviews \ct{22,23}):
\be
              \alpha_A^{-1}(M_{crit}) = 3{\tilde \alpha}_6^{crit}(M_{crit})
                        \approx 3\cdot 1.41\approx 4.23.          \lb{50}
\ee
At the phase transition point $B$, where we have  the breakdown of 
$[\widetilde{E}_6]^3$ into the confinement phase, the one-family value 
${\tilde \alpha}_6^{crit}(\widetilde{M}_{crit})\approx 0.71$ gives the 
following result:
\be
        \alpha_B^{-1}\left(\widetilde{M}_{crit}\right)\approx 3\cdot 0.71
                       \approx 2.13.                        \lb{51} 
\ee
The values (\ref{50}) and (\ref{51}) were used for the construction of the 
curve $AB$ presented in Fig.~\ref{f2}. The point $C$ corresponds to the Planck
scale and $\alpha^{-1}(M_{Pl})=3$, according to Eq.~(\ref{fam}).

The condensation of spreons at the Planck scale predicts the existence 
of a second minimum of the effective potential $V_{eff}(\mu)$ at the scale 
$\mu=M_{Pl}$. The behaviour of this potential and its relation with the 
Multiple Point Principle (see Refs.~\ct{31} and review \ct{23}) will be  
considered in our future investigations.

The dotted curve in Fig.~\ref{f2} describes the running of $\alpha^{-1}(\mu)$
for monopolic ``quark-leptons'' created by preons which are bound by 
supersymmetric hyper-electric non-Abelian flux tubes. We assume that such 
monopoles do not 
exist in our world. However, they can play an essential role in vacuum 
polarization \ct{38} and in the solution of the Cosmological Constant
problem \ct{23,38a}.

\section{The consequences of the preonic model }

In this Section, using our preonic model, we give a simple explanation of
why quarks and leptons of three SM generations have such 
different masses. We show that the hierarchy of SM masses is connected
with the string construction of preon bound states.

We also present a description of form factors related with the compositeness 
of quark-leptons and bosons, briefly discussing a problem of 
compactification in the framework of higher dimensional theories (see 
Ref.~\ct{10}). 

\subsection{Yukawa couplings and quark-lepton masses of three generations}

New preon-antipreon pairs can be generated in the flux tubes between preons. Our 
assumption is that the mechanism of preon-antipreon pair production in these
tubes is similar to that of $e^+e^-$ pair production in a uniform constant
electric field considered in one space dimension \ct{39}.

Almost thirty years ago J.~Schwinger \ct{39} obtained the following expression for
the rate per unit volume that a $e^+e^-$ pair will be created in a constant 
electric field of strength $\cal E$:
\be
           W = \frac {(e{\cal E})^2}{4\pi^3} \sum_{n=1}^{\infty}\frac 1{n^2}
                     \exp\left(-\frac{\pi m_e^2n}{e{\cal E}}\right),           \lb{52}
\ee
where $e$ is the electron charge and $m_e$ is its mass.

A number of authors have applied this result to hadronic production, 
considering the creation of quark-antiquark pairs in QCD tubes of colour
flux (see, for example, Refs.~\ct{40,41,42}).

Considering that the probability $W_k$ of one preon-antipreon pair
production in a unit space-time volume of the $k$-string tube 
with tension (\ref{39a}) is given by Schwinger's expression:
\be
              W_k = \frac{T_k^2}{4\pi^3}\exp\left(-\frac{\pi M_k^2}{T_k}\right),
            \quad  {\rm where}\quad T_k=kT_0,\quad k=1,2,3,                  \lb{53}
\ee
we assume that the Yukawa couplings $Y_k$ for the three generations of quarks 
($t,c$ and $u$) are proportional to the square root of the probabilities $W_k$. 
Then we obtain the following ratios:
\be
       Y_t : Y_c : Y_u :: W_1^{1/2} : W_2^{1/2} : W_3^{1/2}.              \lb{54}
\ee
In Eq.~(\ref{53}) the values $M_k$ are the constituent masses of preons 
produced in the $k$-strings. The Yukawa couplings $Y_k$ are proportional to 
the masses of the $t,c,u$ quarks, and so we have:
\be
       m_t : m_c : m_u :: \frac{T_0}{2\pi^{3/2}}\exp\left(-\frac{\pi M_1^2}{2T_0}\right) :
           \frac{2T_0}{2\pi^{3/2}}\exp\left(-\frac{\pi M_2^2}{4T_0}\right) :
            \frac{3T_0}{2\pi^{3/2}}\exp\left(-\frac{\pi M_3^2}{6T_0}\right).      \lb{55}
\ee
Assuming that $M_k$ is proportional to $k$:
\be
                  M_k =kM_0,                    \lb{56}
\ee
we get the following result:
\be
 m_t : m_c : m_u = m_0\left(w : 2w^2 : 3w^3\right),      \lb{57}
\ee
where
\be
             w = \exp\left(-\frac{\pi M_0^2}{2T_0}\right),            \lb{58}
\ee
and $m_0$ is a mass parameter.
 
For the parameter value 
\be
                w=w_1\approx 2.9\cdot 10^{-3}                              \lb{59}
\ee 
we obtain the following values for the masses of the $t,c,u$-quarks:
\be
         m_t\approx 173\,\,{\rm GeV},\quad m_c\approx 1\,\,{\rm GeV}
         \quad {\rm and}\quad m_u\approx 4\,\,{\rm MeV}.   \lb{60}
\ee                                            
For $b,s,d$-quarks we have different parameters $M_0$, $m_0$ and $w=w_2$.
The value:
\be
                   w_2\approx 1.7\cdot 10^{-2}                \lb{61}
\ee
gives the following masses for the $b,s,d$-quarks: 
\be
         m_b\approx 4\,\,{\rm GeV},\quad m_s\approx 140\,\,{\rm MeV}
         \quad {\rm and}\quad m_d\approx 4\,\,{\rm MeV}.   \lb{62}
\ee       
The results (\ref{60}) and (\ref{62}) are in agreement with experimentally 
established quark masses published in Ref.~\ct{43}:
\be
         m_t\approx 174\pm 5.1\,\,{\rm GeV},\quad m_c\approx 1.15  
\,\,{\rm to}\,\,1.35\,\,{\rm GeV}
         \quad {\rm and}\quad m_u\approx 1.5 \,\,{\rm to}\,\,4 
                 \,\,{\rm MeV}.   \lb{63}
\ee          
and
\be
         m_b \approx 4.1\,\,{\rm to}\,\,4.9
    \,\,{\rm GeV},\quad m_s\approx 80\,\,{\rm to}\,\,130
                  \,\,{\rm MeV}
         \quad {\rm and}\quad m_d\approx 4\,\,{\rm to}\,\,8
         \,\,{\rm MeV}.   \lb{64}
\ee      
The value of the $w$-parameter:
\be
                   w_3\approx 2.5\cdot 10^{-2}                \lb{65}
\ee
leads to the following values for the masses of the $\tau,\mu$-leptons and 
electron: 
\be
   m_{\tau}\approx 2\,\,{\rm GeV},\quad m_{\mu}\approx 100\,\,{\rm MeV}
    \quad {\rm and}\quad m_e\approx 3.5 \,\,{\rm MeV}, \lb{66}
\ee              
which are comparable with the results of experiment \ct{43}:  
\be
   m_{\tau}\approx 1.777\,\,{\rm GeV},\quad m_{\mu}\approx 105.66\,\,
{\rm MeV}\quad {\rm and}\quad m_e\approx 0.51\,\,{\rm MeV}. \lb{67}
\ee      
We see that our preonic model explains the hierarchy of masses
existing in the SM.

\subsection{Form factors as an indication  of the compositeness of quark-leptons }

Non-point-like behaviour of the fundamental quarks and leptons is related
with the appearance of form factors describing the dependence of cross 
sections on 4-momentum squared $q^2$ for different elementary particle 
processes; for example, in the reactions:
$$
    e^+e^- \to e^+e^-,\,\,\mu^+\mu^-,\,\,\tau^+\tau^-,\,\,\gamma\gamma,
$$ or
$$ 
             e^+e^- \to\,\,{\rm hadrons},
$$ 
or in deep inelastic lepton scattering experiments, etc.

Considering the form factor $F(q^2)$ as a power series in $q^2$, we find:
\be
   F(q^2) = 1 + \frac{q^2}{\Lambda^2} + O\left(\frac 1{\Lambda^4}\right),
               \quad  {\rm where} \quad q^2 \ll \Lambda^2.  
                                                            \lb{68}
\ee
If $\Lambda$ $(\Lambda=\Lambda_4$, see Section 4)
is the energy scale associated with the quark-lepton 
compositeness scale $\Lambda_s$ of the preon interaction, then our 
model predicts:
\be
      \Lambda \sim \Lambda_s\sim 10^{18}\,\,{\rm GeV}. \lb{69}
\ee
Of course, such a scale is not experimentally available even for 
future high energy colliders.

From a phenomenological point of view it is quite important to understand 
whether the compactification scale lowers down  $\Lambda$ in Eq.~(\ref{68})
to energies accessible for future accelerators.

Proposing the presence of extra space-time dimensions at small distances
\ct{10} we have a hope that the radius $R_C$ of compactification is larger than 
the radius $R_s$ of compositeness:
\be
                  R_C  \gg  R_s,      \lb{70}
\ee
and
\be
 \Lambda \sim \Lambda_C \ll 10^{18}\,\,{\rm GeV}. \lb{71}
\ee
However, this problem leads to more detailed investigations 
in higher dimensional theories.

\section{Conclusions}

\begin{itemize}
\item[{\bf i.}] In the present paper, starting with the idea that the most realistic 
model based on superstring theory, which leads to the unification of all  
fundamental interactions including gravity, is the ``heterotic'' string-derived 
flipped model, we have assumed that at high energies $\mu > 10^{16}$ GeV
there exists the following chain of flipped models:
$$ SU(3)_C\times SU(2)_L\times U(1)_Y \to
 SU(3)_C\times SU(2)_L\times U(1)_Z \times U(1)_X \to 
$$ $$ SU(5)\times U(1)_X \to  SU(5)\times U(1)_{Z1}
\times U(1)_{X1} \to SO(10) \times U(1)_{X1} \to E_6,$$
ending in the flipped $E_6$ gauge symmetry group (see Fig.~\ref{f1}). 
We have chosen such Higgs
boson contents of the $SU(5)$ and $SO(10)$ gauge groups, which give the final 
flipped $E_6$ unification at the scale $\sim 10^{18}$ GeV and the decreased 
running of $\alpha^{-1}(\mu)$ near the Planck scale given by Fig.~\ref{f2}.

\item[{\bf ii.}] The final $E_6$ unification assumes the existence 
of three 27-plets of the flipped $E_6$ gauge symmetry group and we have 
considered the content of the flipped 27-plet.

\item[{\bf iii.}] The supersymmetric $E_6$ content in five space-time dimensions was 
presented in Sect.~3. The problem of compactification was briefly discussed.

\item[{\bf iv.}] Suggesting an ${\rm\bf N}=1$ supersymmetric 
$E_6\times \widetilde{E_6}$ preonic 
model of composite quark-leptons and bosons, we have assumed that preons are 
dyons confined by hyper-magnetic strings in the region of energies 
$\mu \lesssim M_{Pl}$. This approach is an extension of the old idea by
J.~Pati \ct{5} to use the strong magnetic forces which may bind preons-dyons 
in composite particles -- quark-leptons and bosons. Our 
model is based on the recent theory of composite non-Abelian flux tubes 
in SQCD, which was developed in Refs.~\ct{8a,8b,8c,8d,8e,8f,8p,8q,8r,8s}.

\item[{\bf v.}] Considering the breakdown of $E_6$ (or $\widetilde{E_6}$) 
at the Planck scale
into the $SU(6)\times U(1)$ (or $\widetilde{SU(6)}\times \widetilde{U(1)}$)
gauge group, we have shown that six types of 
$k$-strings -- composite ${\rm\bf N}=1$ supersymmetric non-Abelian flux tubes --
are created by the condensation of spreons near the Planck scale.

\item[{\bf vi.}] It was shown that the six types of strings-tubes, having six fluxes
quantized according to the $Z_6$ center group of
$SU(6)$:
$$
      \Phi_n = n\Phi_0, \quad n=\pm 1,\pm 2,\pm 3,
$$
produce three (and only three) generations of composite quark-leptons and 
bosons. We have obtained a specific type of ``horizontal symmetry'' explaining 
flavor.

\item[{\bf vii.}] It was shown that in the present model preonic strings are very thin,
with radius
$$
          R_{str}\sim 10^{-18}\,\,{\rm GeV}^{-1},
$$
and their tension is enormously large:
$$  
                   T\sim  10^{38}\,\,{\rm GeV}^2.
$$
These strings are similar to the superstrings of Superstring theory.
 
\item[{\bf viii.}] The model predicts the existence of three families of 27-plets and
also gauge bosons $A^i_{\mu}$ (with $i=1,2,3$) belonging to the 78-plets of 
$E_6$. Then near the Planck scale we have the Family 
replicated gauge group of symmetry $[E_6]^3$ for quark-leptons. In the present paper 
we have assumed that the breakdown $[E_6]^3\to E_6$ occurs near the Planck scale 
leading to the $E_6$ unification at the scale $\sim 10^{18}$ GeV.

\item[{\bf ix.}] We have considered the condensation of spreons near the Planck scale, 
which gives phase transitions at the scales $M_{crit}$ and $\widetilde{M}_{crit}$ shown
in Fig.~\ref{f2}. The points $A$ (or $B$) of Fig.~\ref{f2} correspond to the breakdown
of $E_6$ (or $\widetilde{E}_6$) for preons:
$$   E_6 \to SU(6)\times U(1) \quad \left({\rm or} \quad 
                  \widetilde{E}_6 \to \widetilde {SU(6)}\times 
                                        \widetilde {U(1)}\right).
$$
We have investigated the idea that hyper-magnetic strings are produced and exist
at $\mu \le \widetilde{M}_{crit}$, and hyper-electric strings are created and
exist at $\mu \ge M_{crit}$ (the positions of $M_{crit}$ and 
$\widetilde{M}_{crit}$ are shown in Fig.~\ref{f2} by the points $A$ and $B$, respectively).
As a result, in our world we have quark-leptons and gauge bosons $A_{\mu}$ 
(in the region of energies $\mu \lesssim M_{Pl}$), while monopolic 
``quark-leptons" and dual gauge fields $\widetilde{A}_{\mu}$ exist in the region
$\mu \gtrsim M_{Pl}$.
We have calculated the critical values of the gauge coupling constants
at the points $A$, $B$: 
$$
 \alpha^{-1}(M_{crit}) \approx 4.23, \quad {\rm and} \quad
\alpha^{-1}\left(\widetilde{M}_{crit}\right) \approx 2.13.
$$ 

\item[{\bf x.}] Assuming the existence of the three types of hyper-magnetic fluxes 
and using Schwinger's formula, we have given an explanation of the hierarchies of 
masses established in the SM. The following values of the masses were obtained 
in our preonic model: 
$$ m_t\approx 173\,\,{\rm GeV},\quad m_c\approx 1\,\,{\rm GeV}
         \quad {\rm and}\quad m_u\approx 4\,\,{\rm MeV}, $$
$$
  m_b \approx 4\,\,{\rm GeV},\quad m_s\approx 140\,\,{\rm MeV}
         \quad {\rm and}\quad m_d\approx 4\,\,{\rm MeV}, $$
$$
 m_{\tau}\approx 2\,\,{\rm GeV},\quad m_{\mu}\approx 100\,\,{\rm MeV}
    \quad {\rm and}\quad m_e\approx 3.5\,\,{\rm MeV}. 
$$ 
They are comparable with the experimentally known results published in 
Ref.~\ct{43}.

\item[{\bf xi.}] We have considered form factors describing the compositeness of 
quark-leptons. In our preonic model the scale $\Lambda_s$ 
of the quark-lepton compositeness is very large:
$$
\Lambda_s \sim 10^{18}\,\,{\rm GeV},$$
which is not accessible even for future high energy colliders.
We have concluded that only the compactification procedure of extra space-time 
dimensions can help to lower the scale $\Lambda_s$. Then quark-lepton form 
factors might be experimentally observed.

\end{itemize}

\section*{Acknowledgements}

C.R.D. thanks Prof.~J.~Pati a lot for interesting discussions during the 
Conference WHEPP-9 (Bhubaneswar, India, January, 2006). 

L.L. sincerely thanks the Institute of Mathematical Sciences (Chennai, India) 
and personally the Director of IMSc Prof.~R.~Balasubramanian and  
Prof.~N.D.~Hari Dass for the wonderful hospitality and financial support.

The authors also deeply thank Prof.~J.L.~Chkareuli,
Prof.~F.R.~Klinkhamer and Prof.~G.~Rajasekaran for fruitful discussions and 
advice. 

We are thankful of Prof.~C.D.~Froggatt for his great help.

\section*{Appendix A: Compactification of extra dimensions 
in preonic model}
\setcounter{equation}{0}
\renewcommand{\theequation}{A.\arabic{equation}}

Compactifying the extra fifth dimension $x^4$ in Eq.~(\ref{17}) on a circle
of radius $R_C$, the authors of Ref.~\ct{10} impose the Scherk-Schwarz 
\ct{14} boundary conditions to the preonic superfields:
\bea
   P\left(x^m, x^4+2\pi R_C, \theta\right) &=& e^{i2\pi q_P}P\left(x^m, x^4, e^{i\pi
(q_P+q_{\bar P})}\theta\right),                  
\nonumber\\
   P^{\rm\bf c}\left(x^m, x^4+2\pi R_C, \theta\right) &=& 
e^{i2\pi q_{\bar P}}P^{\rm\bf c}\left(x^m, x^4, e^{i\pi
(q_P+q_{\bar P})}\theta\right),                  
\nonumber\\
   P_s\left(x^m, x^4+2\pi R_C, \theta\right) &=& e^{i2\pi q_s}P_s\left(x^m, x^4, e^{i\pi
(q_s+q_{\bar s})}\theta\right),                  
\nonumber\\
   P^{\rm\bf c}_s\left(x^m, x^4+2\pi R_C, \theta\right) &=& e^{i2\pi q_{\bar s}}P^{\rm\bf c}_s
\left(x^m, x^4, e^{i\pi(q_s+q_{\bar s})}\theta\right),                  \lb{a1}
\eea
where $q_P$, $q_{\bar P}$ and  $q_s$, $q_{\bar s}$ are the $R$ charges of preons 
$P$, $P^{\rm\bf c}$ and $P_s$, $P^{\rm\bf c}_s$, respectively.
 The following condition was obtained in Ref.~\ct{10}:
 \be   
      q_P+q_{\bar P} = q_s+q_{\bar s}.              \lb{a2}
\ee
Then the $R$ charges for the composite states (\ref{21})--(\ref{24}) are:
\be Q\sim q_P + q_{\bar s},\quad \bar Q\sim  q_{\bar P} + q_{s},
  \quad M \sim  q_P + q_{\bar P}, \quad S\sim  q_s + q_{\bar s}.    \lb{a3}
\ee
As a result of compactification, all fermionic preons are massive in $4D$ 
space-time. Supersymmetry is broken by the boundary conditions (\ref{a1}).

\clearpage\newpage
\bfi
\centering
\includegraphics[height=163mm,keepaspectratio=true,angle=-90]{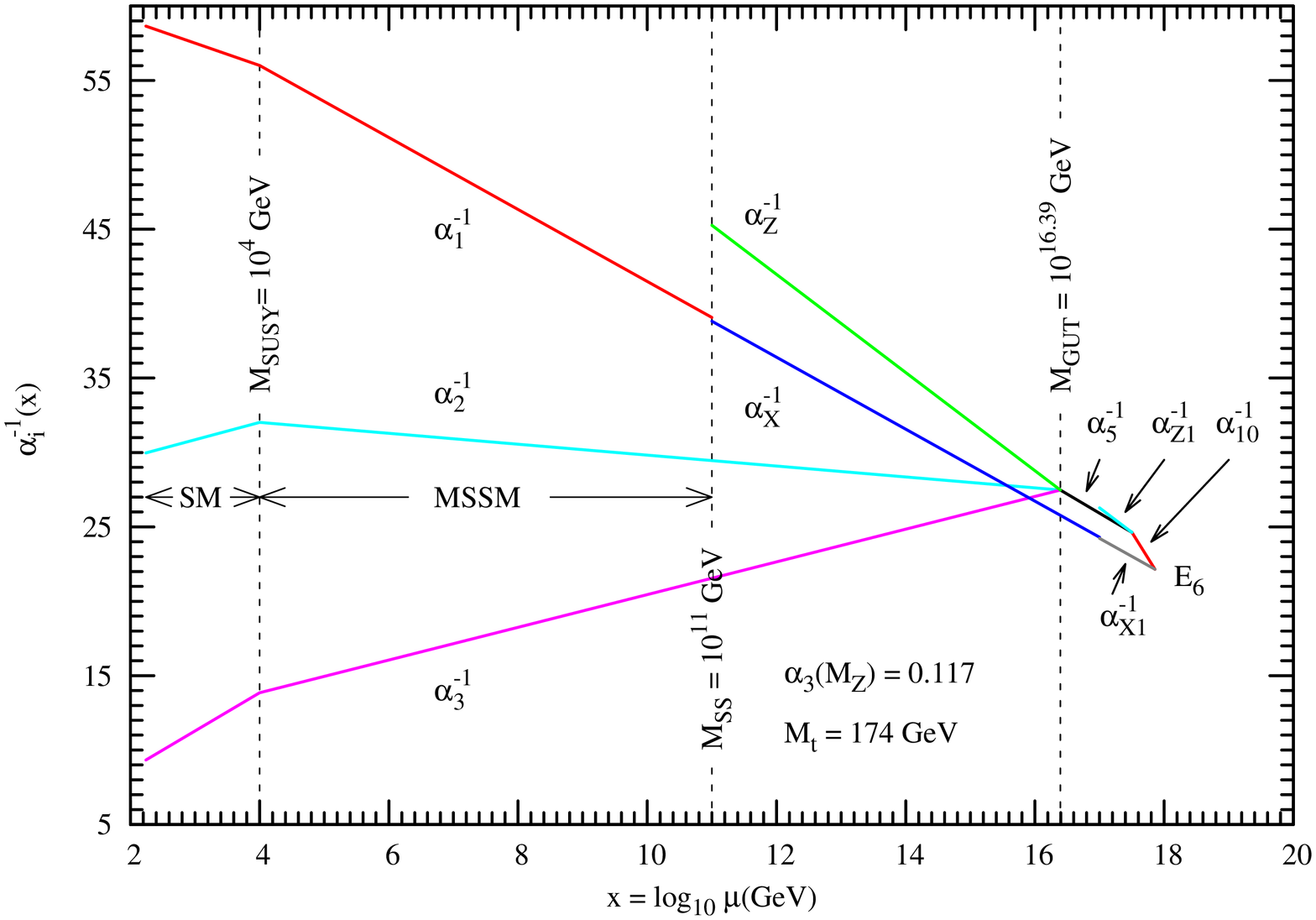}
\caption
{This figure presents the running of the inverse gauge coupling constants
$\alpha_i^{-1}(x)$ for $i=1,2,3,X,Z,X1,Z1,5,10$ of the chain
$ SU(3)_C\times SU(2)_L\times U(1)_Y \to
 SU(3)_C\times SU(2)_L\times U(1)_Z \times U(1)_X \to 
SU(5)\times U(1)_X \to  SU(5)\times U(1)_{Z1}
\times U(1)_{X1} \to SO(10) \times U(1)_{X1} \to E_6
$, corresponding to the breakdown of the flipped $SU(5)$ to the supersymmetric 
(MSSM) $SU(3)_C\times SU(2)_L\times U(1)_Z\times U(1)_X$ gauge symmetry group 
with Higgs bosons belonging to the $5_h + \bar 5_h$, $10_H + \ov {10}_H$,  
24-dimensional adjoint $A$ and higher representations of the flipped $SU(5)$.
The final unification group is flipped $E_6$ at the supersuperGUT scale 
$ M_{SSG}\approx 10^{18}$ GeV with 
$\alpha^{-1}(M_{SSG})\approx 22$ for $M_{SUSY}=10$ TeV and 
seesaw scale $M_{SS}=10^{11}$ GeV.}
\lb{f1}
\efi

\clearpage\newpage
\bfi
\centering
\includegraphics[height=163mm,keepaspectratio=true,angle=-90]{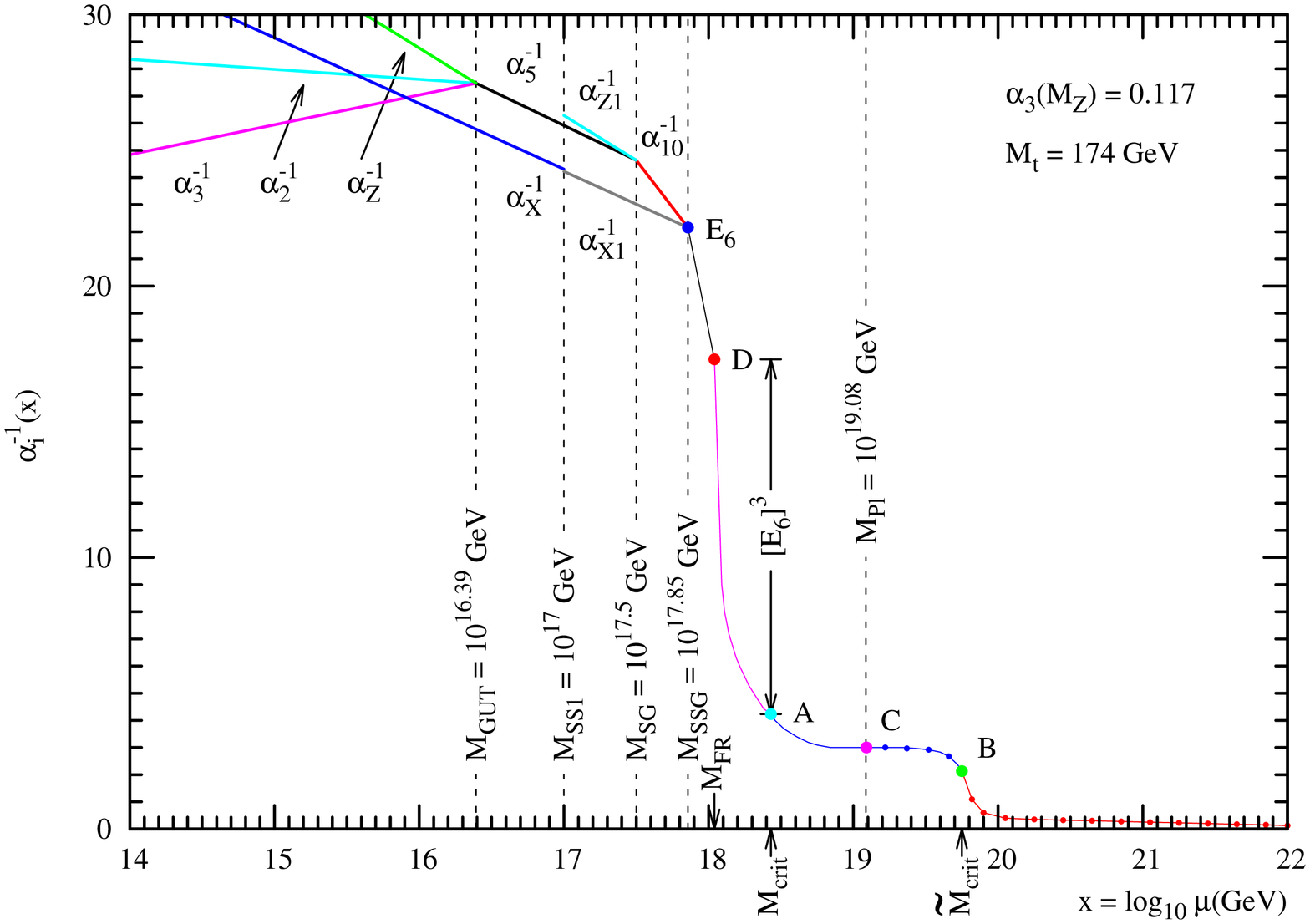}
\caption
{The figure provides a qualitative description of the running of $\alpha^{-1}(x)$ 
near the Planck scale predicted by the present preonic model. The region 
$AD$ corresponds to the Family replicated gauge group of symmetry $[E_6]^3$,
which arises at the scale $M_{FR}$ given by point $D$.
The point $A$ at the scale of energy $\mu=M_{crit}$ indicates that hyper-electric
preonic strings exist for $\mu \ge M_{crit}$. The point $B$ corresponds to 
the scale $\mu = \widetilde{M}_{crit}$ and indicates that hyper-magnetic 
preonic strings exist for  $\mu\le \widetilde{M}_{crit}$. The curve $AB$
corresponds to the region of energies where spreons are condensed near the
Planck scale, giving both hyper-electric and hyper-magnetic, preonic strings.
For  $\mu \ge \widetilde{M}_{crit}$ we have the running of $\alpha^{-1}(x)$
for monopolic ``quark-leptons". The point $C$ corresponds to the Planck scale
and gives $\alpha^{-1}(M_{Pl})=3$.}
\lb{f2}
\efi

\clearpage\newpage

\bfi
\centering
\includegraphics[height=62mm,keepaspectratio=true]{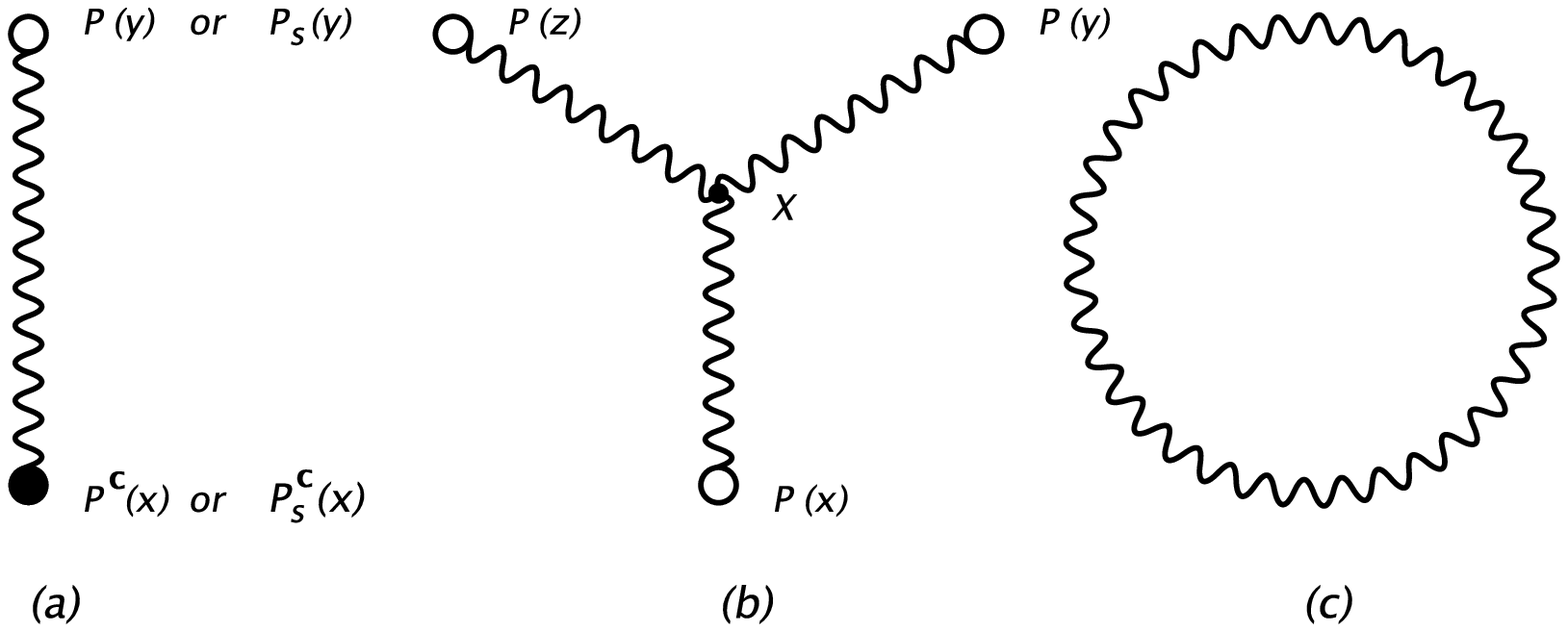}
\caption{Preons are bound by hyper-magnetic strings: (a,b) correspond to
the string configurations of composite particles belonging to the
27-plet of the flipped $E_6$ gauge symmetry group; (c) represents a 
closed string describing a graviton.}
\lb{f3}
\efi

\clearpage\newpage
\bfi
\centering
\includegraphics[height=59mm,keepaspectratio=true]{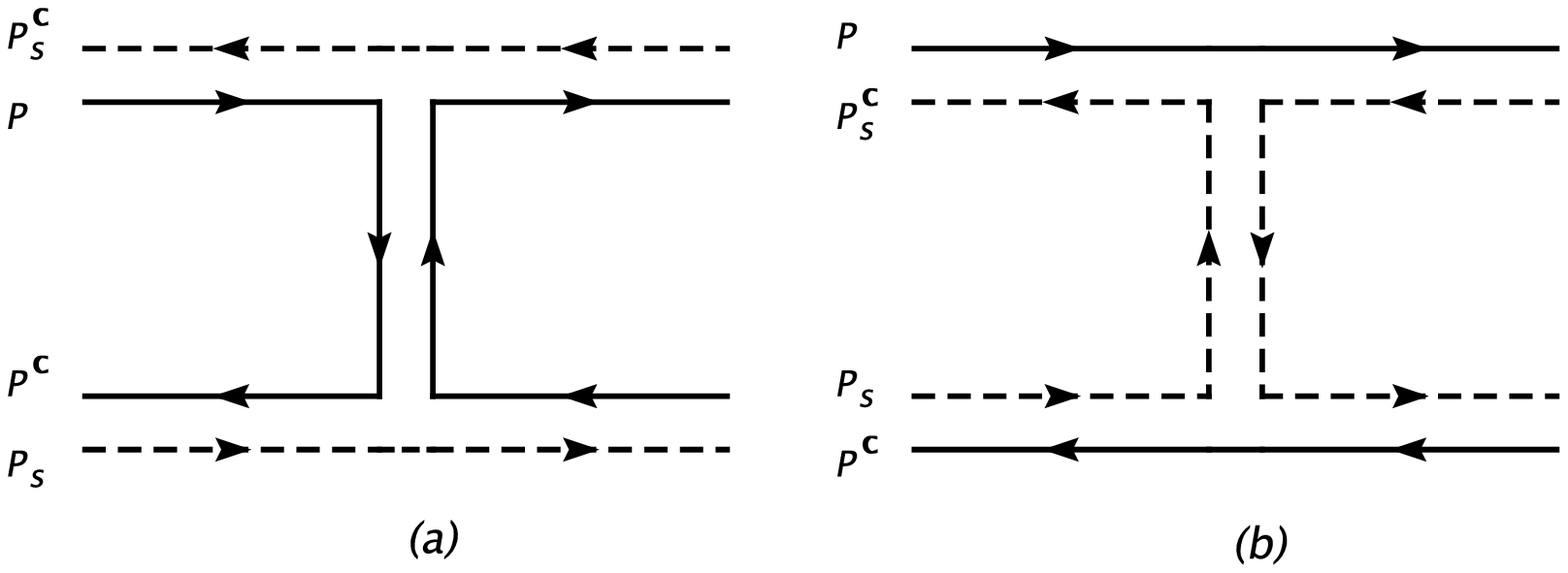}
\caption{Vector gauge bosons belonging to the 78 representation of the flipped $E_6$
and Higgs scalars -- singlets of $E_6$ -- are composite objects created 
(a) by fermionic preons $P,P^{\rm\bf c}$ and (b) by scalar preons $P_s,P^{\rm\bf c}_s$.
Both of them are confined by hyper-magnetic strings.}
\lb{f4}
\efi


\begin{thebibliography}{99}
\bibitem{1}
J.~Schwarz, Phys.Rept. {\bf 89}, 223 (1982); M.B.~Green, Surv.High.En.Phys. 
{\bf 3}, 127 (1984).
\bibitem{2}
D.J.~Gross, J.A.~Harvey, E.~Martinec and R.~Rohm, Phys.Rev.Lett. {\bf 54}, 502
(1985); Nucl.Phys. {\bf B256}, 253 (1985); ibid., {\bf B267}, 75 (1986).
\bibitem{3}
P.~Candelas, G.T.~Horowitz, A.~Strominger and E.~Witten, Nucl.Phys. {\bf B258}, 
46 (1985).
\bibitem{4}
M.B.~Green and J.H.~Schwarz, Phys.Lett. {\bf B149}, 117 (1984); ibid., 
{\bf B151}, 21 (1985). 
\bibitem{5}
J.C.~Pati, Phys.Lett. {\bf B98}, 40 (1981).
\bibitem{6} 
J.C.~Pati, A.~Salam and J.~Strathdee, Nucl.Phys. {\bf B185}, 416 (1981).
\bibitem{6a}
P.~Senjanovic, Fizika (Zegreb) {\bf 15}, 165 (1983);
M.~Blagojevic, S.~Meljanac and P.~Senjanovic, Phys.Lett. {\bf B131}, 111 (1983).
\bibitem{6b}
Dan-di Wu, Phys.Rev. {\bf D34}, 280 (1986).
\bibitem{7}
M.~Blagojevic and P.~Senjanovic, Phys.Rept. {\bf 157}, 233 (1988).
\bibitem{8a}
M.~Shifman and A.~Yung, Phys.Rev. {\bf D72}, 085017 (2005), arXiv: hep-th/0501211;
ibid., {\bf D70}, 045004 (2004), arXiv: hep-th/0403149; ibid., {\bf D70}, 
025013 (2004), arXiv: hep-th/0312257.
\bibitem{8b}
A.~Gorsky, M.~Shifman and A.~Yung, Phys.Rev. {\bf D71}, 045010 (2005); 
arXiv: hep-th/0412082.
\bibitem{8c}
R.~Auzzi, M.~Shifman and A.~Yung, Phys.Rev. {\bf D73}, 105012 (2006); 
arXiv: hep-th/0511150.
\bibitem{8d}
V.~Markov, A.~Marshakov and A.~Yung, Nucl.Phys. {\bf B709}, 267 (2005);
arXiv: hep-th/0408235.
\bibitem{8e}
M.A.C.~Kneipp, Phys.Rev. {\bf D69}, 045007 (2004), arXiv: hep-th/0308086;
M.A.C.~Kneipp and P.~Brockill, Phys.Rev. {\bf D64}, 125012 (2001), 
arXiv: hep-th/0104171. 
\bibitem{8f}
D.~Tong, Phys.Rev. {\bf D69}, 065003 (2004), arXiv: hep-th/0307302;
A.~Hanany and D.~Tong, JHEP {\bf 0404}, 066 (2004) arXiv: hep-th/0403158; 
ibid.,  {\bf 0307}, 037 (2003), arXiv: hep-th/0306150.
\bibitem{8p}
M.~Eto, Y.~Isozumi, M.~Nitta, K.~Ohashi and N.~Sakai,
Phys.Rev.Lett. {\bf 96}, 161601 (2006); 
arXiv: hep-th/0511088.
\bibitem{8q}
Y.~Isozumi, M.~Nitta, K.~Ohashi and N.~Sakai, Phys.Rev. {\bf D71}, 065018 (2005), 
arXiv: hep-th/0405129;
M.~Eto, T.~Fujimori, Y.~Isozumi, M.~Nitta, K.~Ohashi, K.~Ohta and N.~Sakai, 
Phys.Rev. {\bf D73}, 085008 (2006), arXiv: hep-th/0601181;
M.~Eto, Y.~Isozumi, M.~Nitta, K.~Ohashi and N.~Sakai,
Phys.Rev. {\bf D73}, 125008 (2006), arXiv: hep-th/0602289.
\bibitem{8r} M.~Eto, M.~Nitta and N.~Sakai,
Nucl.Phys. {\bf B701}, 247 (2004); arXiv: hep-th/0405161.
\bibitem{8s} 
M.~Eto, Y.~Isozumi, M.~Nitta, K.~Ohashi and N.~Sakai, J.Phys. {\bf A39}, R315 (2006);
arXiv: hep-th/0602170.
\bibitem{9a}
G.S.~Iroshnikov, {\it World -- sheet instantons in the theory of the QCD 
strings}; arXiv: hep-th/0508129.
\bibitem{9}
S.~Sen, Phys.Rev. {\bf D37}, 1296 (1988).
\bibitem{10}
M.~Chaichian, J.L.~Chkareuli and A.~Kobakhidze, Phys.Rev. {\bf D66}, 095013 (2002);
arXiv: hep-ph/0108131.
\bibitem{10a}
H.~Harari, Phys.Rept. {\bf 104}, 159 (1984).
\bibitem{10b}
R.E.~Marshak, Phys.Rept. {\bf 137}, 73 (1986).
\bibitem{10c}
R.R.~Volkas and R.C.~Joshi, Phys.Rept. {\bf 159}, 303 (1988).
\bibitem{10d}
L.B.~Okun, {\it Leptons and  Quarks}
(North-Holland Physics Publishing, Amsterdam, 1982).
\bibitem{10e}
L.~Lyons, Prog.Part.Nucl.Phys. {\bf 10}, 227 (1983);
ibid., {\bf 7}, 157 (1981).
\bibitem{10x}
J.L.~Chkareuli, {\it Composite quarks and leptons: from SU(5) to SU(8)},
in: Proceedings of 
   ``Quarks-82'', Sukhumi, USSR, 5-7 May, 1982 (Institute of 
   Nuclear Research, USSR Academy of Sciences, 
   Moscow, 1983), p.149.
\bibitem{10f}
H.~Stremnitzer, {\it Composite models and SUSY}, in:
Proceedings to the Workshop at Bled, Slovenia, 29 June -- 9 July, 1998: 
``What comes beyond the Standard Model''
(DMFA, Ljubljana, Slovenia, 1999), p.45; arXiv: hep-ph/9905357.
\bibitem{10g}
J.J.~Dugne, S.~Fredriksson, J.~Hansson and E.~Predazzi,
{\it  Preon trinity: A New model of leptons and quarks}, in: 
``Beyond the desert 1999'', Tegernsee, Germany, 6-12 June, 1999,
Ed. H.V. Klapdor-Kleingrothaus and I. Krivosheina
(Institute of Physics Publishing, Bristol and Philadelphia, 2000), p.285;
arXiv: hep-ph/9909569.
\bibitem{10k}
S.~Fredriksson, {\it Preon prophecies by the Standard Model},
in: ``Beyond the desert 2003'', Schloss Ringberg, Bavaria, Germany, 
9-14 June, 2003, Ed. H.V. Klapdor-Kleingrothaus
(Springer, Berlin, 2004), p.211; 
arXiv: hep-ph/0309213.
\bibitem{11}
C.R.~Das and L.V.~Laperashvili, {\it Seesaw scales and the steps from the 
Standard Model towards superstring-inspired flipped $E_6$}; 
arXiv: hep-ph/0604052.
\bibitem{11aa}
V.~Barger, N.G.~Deshpande, R.J.N.~Phillips and K.~Whisnant, 
Phys.Rev. {\bf D33}, 1912 (1986).
\bibitem{11aaa}
J.L.~Hewett and T.G.~Rizzo, Phys.Rept. {\bf 183}, 193 (1989).
\bibitem{11a}
B.~Stech and Z.~Tavartkiladze, Phys.Rev. {\bf D70}, 035002 (2004); 
arXiv: hep-ph/0311161.
\bibitem{11b}
S.F.~King, S.~Moretti and R.~Nevzorov, Phys.Lett. {\bf B634}, 278 (2006), 
arXiv: hep-ph/0511256; arXiv: hep-ph/0601269. 
\bibitem{12}
C.R.~Das, C.D.~Froggatt, L.V.~Laperashvili and H.B.~Nielsen, Mod.Phys.Lett. 
{\bf A21}, 1151 (2006); arXiv: hep-ph/0507182.
\bibitem{13}
H.~Georgi and S.L.~Glashow, Phys.Rev.Lett. {\bf 32}, 438 (1974).
\bibitem{14}
J.~Scherk and J.H.~Schwarz, Phys.Lett. {\bf B82}, 60 (1979).
\bibitem{15}
N.~Arkani-Hamed, T.~Gregoire and J.~Wacker, JHEP {\bf 0203}, 055 (2002); 
arXiv: hep-th/0101233; 
D.~Marti and A.~Pomarol, Phys.Rev. {\bf D64}, 105025 (2001); arXiv: hep-th/0106256.
\bibitem{17}
A.A.~Abrikosov, Sov.Phys.JETP {\bf 5}, 1174 (1957) [Zh.Eksp.Teor.Fiz. {\bf 32}, 1442 (1957)].
\bibitem{18}
H.B.~Nielsen and P.~Olesen, Nucl.Phys. {\bf B61}, 45 (1973).
\bibitem{16}
S.~Cecotti, J.P.~Derendinger, S.~Ferrara and L.~Girardello, Phys.Lett. {\bf 
B156}, 318 (1985).
\bibitem{16a}
J.L.~Chkareuli, JETP Lett. {\bf 32}, 671 (1980) [Pisma Zh.Eksp.Teor.Fiz. {\bf 32}, 684 (1980)]; 
ibid., {\bf 34}, 453 (1981) [Pisma Zh.Eksp.Teor.Fiz. {\bf 34}, 474 (1981)];
ibid., {\bf 36}, 493 (1983) [Pisma Zh.Eksp.Teor.Fiz. {\bf 36}, 407 (1982)].
\bibitem{16b}
Z.G.~Berezhiani and J.L.~Chkareuli, Sov.J.Nucl.Phys. {\bf 37}, 618 (1983) [Yad.
Fiz. {\bf 37}, 1043 (1983)].
\bibitem{16c}
J.L.~Chkareuli, C.D.~Froggatt and H.B.~Nielsen, Nucl.Phys. {\bf B626}, 307 (2002);
arXiv: hep-ph/0109156. 
\bibitem{16d}
T.~Appelquist, Y.~Bai and M.~Piai, Phys.Rev. {\bf D72}, 036005 (2005);
arXiv: hep-ph/0506137.
\bibitem{19a}
Chan Hong-Mo and Tsou Sheung Tsun, Phys.Rev. {\bf D56}, 3646 (1997); 
ibid., {\bf D57}, 2507 (1998).
\bibitem{19b}
Chan Hong-Mo, J.~Faridani and Tsou Sheung Tsun, Phys.Rev. {\bf D51}, 7040 (1995);
ibid., {\bf D52}, 6134 (1995);
ibid., {\bf D53}, 7293 (1996).
\bibitem{19c}
Sheung Tsun Tsou, Int.J.Mod.Phys. {\bf A18S2}, 1 (2003); arXiv: 
hep-th/0110256.
\bibitem{19}
H.B.~Nielsen and Y.~Takanishi, Nucl.Phys. {\bf B604}, 405 (2001), 
arXiv: hep-ph/0011062; 
Phys.Lett. {\bf B507}, 241
(2001), arXiv: hep-ph/0101307; ibid., {\bf B543}, 249 (2002), 
arXiv: hep-ph/0205180.
\bibitem{20}
C.D.~Froggatt, H.B.~Nielsen and Y.~Takanishi, Nucl.Phys. {\bf B631}, 285 (2002); 
arXiv: hep-ph/0201152.
\bibitem{21}
C.D.~Froggatt and H.B.~Nielsen, {\it Trying to understand the Standard
Model parameters}. Invited talk by H.B.~Nielsen at the ``XXXI
ITEP Winter School of Physics'', Moscow, Russia, 18-26 February,
2003, published in Surveys High Energ.Phys. {\bf 18}, 55 (2003);
arXiv: hep-ph/0308144.
\bibitem{22}
C.D.~Froggatt, L.V.~Laperashvili, H.B.~Nielsen and Y.~Takanishi,
{\it Family Replicated Gauge Group Models}, in: Proceedings of the
``Fifth International Conference `Symmetry in Nonlinear Mathematical
Physics'', Kiev, Ukraine, 23-29 June, 2003, Ed. by A.G.~Nikitin,
V.M.~Boyko, R.O.~Popovich and I.A.~Yehorchenko (Institute of Mathematics
of NAS of Ukraine, Kiev, 2004), Vol.50, Part 2, p.737; arXiv:
hep-ph/0309129.
\bibitem{23}
C.R.~Das and L.V.~Laperashvili, Int.J.Mod.Phys. {\bf A20}, 5911 (2005);
arXiv: hep-ph/0503138.
\bibitem{24}
F.S.~Ling and P.~Ramond, Phys.Rev. {\bf D67}, 115010 (2003); 
arXiv: hep-ph/0302264.
\bibitem{25}
S.~Rajpoot, Nucl.Phys.Proc.Suppl. {\bf A51}, 50 (1996).
\bibitem{26}
S.~Dimopoulos, G.~Dvali, R.~Rattazzi and G.F.~Giudice, Nucl.Phys.
{\bf B510}, 12 (1998); arXiv: hep-ph/9705307.
\bibitem{27}
L.V.~Laperashvili, H.B.~Nielsen and D.A.~Ryzhikh, 
Int.J.Mod.Phys. {\bf A18}, 4403 (2003), arXiv: hep-th/0211224;
Phys.At.Nucl. {\bf 65}, 353 (2002) [Yad.Fiz. {\bf 65}, 377 (2002)], arXiv: hep-th/0109023; 
ibid., {\bf 66}, 2070 (2003) [Yad.Fiz. {\bf 66}, 2119 (2003)], arXiv: hep-th/0212213.
\bibitem{28}
L.V.~Laperashvili and D.A.~Ryzhikh, {\it Phase transition in gauge
theories and the Planck scale physics}, in: 
Proceedings ``Bled 2000/2001, What comes beyond the Standard Model'', Ed.
M.~Breskvar et al.
(DMFA, Zaloznistvo, Ljubljana, December, 2002), 
Vol.1, pp.83-100, arXiv: hep-th/0110127; {\it $[SU(5)]^3$ SUSY unification}, 
arXiv: hep-th/0112142.
\bibitem{29}
D.L.~Bennett and H.B.~Nielsen, I.~Picek, Phys.Lett. {\bf B208}, 275
(1988).
\bibitem{30}
C.D.~Froggatt and H.B.~Nielsen, {\it Origin of Symmetries}, (World
Sci., Singapore, 1991).
\bibitem{31}
D.L.~Bennett and H.B.~Nielsen, Int.J.Mod.Phys. {\bf A9}, 5155
(1994), arXiv: hep-ph/9311321; D.L.~Bennett, C.D.~Froggatt and 
H.B.~Nielsen, in: Proceedings of
the ``27th International Conference on High Energy Physics'', Glasgow,
Scotland, 1994, Ed. by P.~Bussey and I.~Knowles (IOP Publishing
Ltd, Bristol, 1995), p.557; {\it Perspectives in Particle Physics '94}, Ed.
by D.~Klabu\u{c}ar, I.~Picek and D.~Tadi\'{c} (World Scientific,
Singapore, 1995), p.255; arXiv: hep-ph/9504294.
\bibitem{32}
L.V.~Laperashvili, Phys.At.Nucl. {\bf 57}, 471 (1994) [Yad.Fiz. {\bf 57}, 501 (1994)]; 
ibid., {\bf 59}, 162 (1996) [Yad.Fiz. {\bf 59}, 172 (1996)].
\bibitem{33}
L.V.~Laperashvili, H.B.~Nielsen and D.A.~Ryzhikh, Int.J.Mod.Phys. {\bf A16}, 
3989 (2001); arXiv: hep-th/0105275.
\bibitem{34}
C.R.~Das, L.V.~Laperashvili and H.B.~Nielsen, {\it Generalized dual symmetry of 
non-Abelian theories and the freezing of $\alpha_s$}, 
to appear in Int.J.Mod.Phys. {\bf A} (2006); arXiv: hep-ph/0511267.
\bibitem{35}
J.~Jersak, T.~Neuhaus and P.M.~Zerwas, Phys.Lett. {\bf B133}, 103 (1983); 
Nucl.Phys. {\bf B251}, 299 (1985).
\bibitem{36}
J.~Jersak, T.~Neuhaus and H.~Pfeiffer, Phys.Rev. {\bf D60}, 054502 (1999); 
arXiv: hep-lat/9903034.
\bibitem{37}   
L.V.~Laperashvili and H.B.~Nielsen, Int.J.Mod.Phys. {\bf A16}, 2365 (2001); 
arXiv: hep-th/0010260. 
\bibitem{37a}
C.R.~Das and L.V.~Laperashvili, {\it Phase transition in the Higgs model of scalar dyons},
to appear in Mod.Phys.Lett. {\bf A} (2006); arXiv: hep-ph/0511067.
\bibitem{38}
F.R.~Klinkhamer and G.E.~Volovik, JETP Lett. {\bf 81}, 551 (2005)
[Pisma Zh.Eksp.Teor.Fiz. {\bf 81}, 683 (2005)]; 
arXiv: hep-ph/0505033.
\bibitem{38a}
C.~Froggatt, L.~Laperashvili, R.~Nevzorov and H.B.~Nielsen,
Phys.Atom.Nucl. {\bf 67}, 582 (2004) [Yad.Fiz. {\bf 67}, 601 (2004)]; arXiv: hep-ph/0310127.
\bibitem{39}
J.~Schwinger, {\it Particles, sources and fields} (Addison-Wesley, 1973).
\bibitem{40}
A.~Casher, H.~Neuberger and S.~Nussinov, Phys.Rev. {\bf D20}, 179 (1979).
\bibitem{41}
E.G.~Gurvich, Phys.Lett. {\bf B87}, 386 (1979).
\bibitem{42}
N.K.~Glendenning and T.~Matsui, Phys.Rev. {\bf D28}, 2890 (1983).
\bibitem{43}
S.~Eidelman et al., Particle Data Group, Phys.Lett. {\bf B592}, 1 (2004).

\end{thebibliography}
\end{document}